\documentclass[aps,prl,twocolumn,showpacs,groupaddress]{revtex4}
\usepackage{amssymb}
\usepackage{graphicx}
\usepackage{dcolumn}
\usepackage{bm}
\usepackage{amsmath}

\begin{document}

\title{Temporal shaping of entangled photons}
\author{Avi Pe'er}
\author{Barak Dayan}
\author{Asher A. Friesem}
\author{Yaron Silberberg}
\affiliation{Department of Physics of Complex Systems, Weizmann
Institute of Science, Rehovot 76100, Israel}

\begin{abstract}

We experimentally demonstrate shaping of the two-photon
wavefunction of entangled photon-pairs, utilizing coherent
pulse-shaping techniques. By performing spectral-phase
manipulations we tailor the two-photon wavefunction exactly like a
coherent ultrashort pulse. To observe the shaping we perform
sum-frequency generation (SFG) with an ultrahigh flux of entangled
photons. At the appropriate conditions, SFG performs as a
coincidence detector with an ultrashort response time ($\sim 100
\: fs$), enabling a direct observation of the two-photon
wavefunction. This property also enables us to demonstrate
background-free, high-visibility two-photon interference
oscillations.
\end{abstract}


\maketitle

Entangled photons generated by parametric down-conversion can
exhibit nonclassical correlations between various physical
properties such as polarization, momentum and energy, and have
been the primary tool in quantum communication science
\cite{Ou@Mandel_PRL_1990,Ekert@Palma_PRL_1992,Mattl@Zeilinger_PRL_1996,
Bouwmeester@Zeilinger_Nature_1997,Jennewein@Zeilinger_PRL_2000}.
While the polarization state of polarization-entangled photons has
been readily manipulated in various quantum-optics experiments
(see, for example
\cite{{Mattl@Zeilinger_PRL_1996},{Bouwmeester@Zeilinger_Nature_1997},{Jennewein@Zeilinger_PRL_2000}}),
no control over their temporal properties has been demonstrated to
date, except the prolongation of their correlation-time by
spectral filtering with narrowband filters
\cite{Ou@Lu_PRL_1999,Bellini@Arecchi_PRL_2003,{Viciani@Bellini_PRA_2004}}.\

Here we demonstrate that the two-photon wave function of entangled
photons can be shaped by spectral-phase manipulations exactly like
classical pulses. The physical reason for this lies in the
coherent spectrum of the two-photon state, which governs the
temporal correlations between the photons in the very same way
that the spectrum of a coherent pulse determines its shape.\

Let us consider the two-photon state $|\Psi\rangle$ created by
type-I, broadband down-conversion of a narrowband pump
\cite{Shaped_entangled_note0}:

\begin{eqnarray} \label{bi}
|\Psi\rangle \propto M | 0 \rangle + \int d \omega \: g(\omega) \:
| \omega \rangle_s | \omega_p-\omega \rangle_i
\end{eqnarray}

\noindent where $M\sim 1$, and the subscripts $s,i,p$ denote the
signal, idler and pump modes, respectively. $g(\omega)$ is
determined by the nonlinear coupling and the phase-matching
conditions in the down-converting crystal, and exhibits a constant
spectral phase for the case of perfect phase matching. At low
photon fluxes, the two-photon wavefunction $\psi(t_s,t_i)$
represents the probability amplitude for the joint detection of
signal and idler photons at times $t_s,t_i$ respectively, and is
defined by:

\begin{eqnarray} \label{two}
\psi(t_s,t_i)=\langle 0 |E^+_s\: E^+_i \: |\Psi\rangle
\end{eqnarray}

Introducing spectral filters $\Phi_s(\omega)$,$\Phi_i(\omega)$ to
the signal and idler modes, and assuming a gaussian spectrum with
a bandwidth $\delta_p$ for the pump, the two-photon wavefunction
can be approximated by
\cite{{Keller@Rubin_PRA_1997},Shaped_entangled_note1}:
\begin{eqnarray}
\label{F} \psi(t_s,t_i) \propto
e^{-\frac{1}{32}\delta_p^2(t_s+t_i)^2} G(t_s-t_i)
\end{eqnarray}
where $G(t)$ is the inverse Fourier transform of
$g(\omega)\Phi_s(\omega)\Phi_i(\omega_p-\omega)$ as a function of
$\omega$. In other words, the two-photon wavefunction behaves like
a coherent pulse with a spectrum $g(\omega)$, that was shaped by a
spectral filter
$\Phi(\omega)=\Phi_s(\omega)\Phi_i(\omega_p-\omega)$. This shaping
can be nonlocal, since both the filtering and the detection of the
signal and the idler photons can be located at different places.
Note that similarly, the two-photon wavefunction of momentum
entangled photons (see \cite{Monken@Padua_PRA_1998}) could be
spatially shaped by phase manipulations in the momentum space. In
collinear down-conversion (as in our case), it is convenient to
define the higher (lower) energy photon as the signal (idler), or
vice versa. In such a case $g(\omega)$ is symmetric about
$\omega_p/2$.\

Although our discussion so far dealt with continuous
down-conversion, shaping of the two-photon wavefunction is
possible with pulsed down-conversion as well. In this case, the
signal and the idler are both pulses with a constant (though
undefined) spectral phase, like the pump. Thus, obviously, each
can be shaped independently by spectral-phase filters, leading to
the same result as in the continuous case
\cite{Shaped_entangled_note2}.\

The most direct way to observe the simultaneous arrival of two
photons is to detect a nonlinear photon-photon interaction between
them. Due to the extremely low efficiencies of nonlinear
interactions with entangled photons, the detection of their
inherent nonclassical correlations has been limited so far to
photoelectric coincidences. Nonetheless, as was recently
pointed-out \cite{Dayan@Silberberg_QPH_2004}, it is possible to
generate surprisingly high fluxes of entangled photons without
losing their nonclassical properties. Since the photons of each
photon-pair are temporally correlated to within a time scale that
is inversely proportional to their bandwidth $\Delta_\textsc{dc}$,
the maximal flux $\Phi_{max}$ at which down-converted light can
still be considered as composed of separated photon-pairs is:
\begin{eqnarray}
\label{LinQSFG_eq1} \Phi_{max}\approx\Delta_\textsc{dc} \ .
\end{eqnarray}

This flux corresponds to a mean spectral photon-density of $n=1$.
For broadband down-conversion, $\Phi_{max}$ can exceed $10^{13}$
photons per second, which corresponds to classically-high power
levels of about $2\:\mu W$. Such ultrahigh fluxes of broadband
entangled photons enable the demonstration of nonlinear
interactions even with interaction efficiencies of the order of
$10^{-9}$, which can be achieved without special enhancement
mechanisms.\

In this work we introduce the use of sum-frequency generation
(SFG) as an ultrafast coincidence detector for photon fluxes that
are below $\Phi_{max}$. The sensitivity of the SFG process to
relative delay between the photons is inversely proportional to
the interaction's bandwidth, which for broadband phase-matching
can be of the order of $\sim 100\:fs$. Such response times are
completely unattainable with current electronic technologies, and
allow a direct observation of the two-photon wavefunction.\

\begin{figure}[t]
\begin{center}
\includegraphics[width=8.6cm] {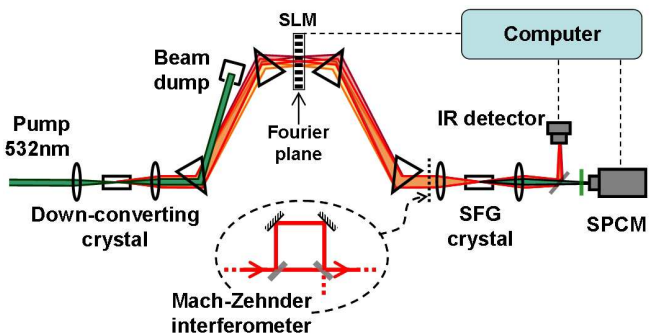}
\caption{\label{fig1} Experimental layout. Entangled photons
down-converted in one crystal are up-converted in the other
crystal to produce the SFG photons. The symmetric imaging
arrangement was designed to have a focus in the mid-way between
the second and third prisms, creating there a spectral
Fourier-plane. A computer-controlled SLM located at this plane was
used to perform spectral-phase manipulations on the entangled
photons. The SFG photons were counted using a single-photon
counting module (SPCM-AQR-15 of $EG\&G$), with a dark-count of
$\sim50\:s^{-1}$. This dark-count was subtracted from all our
measurements, which were performed with integration times of
$5s$-$10s$. The entangled-photons beam was filtered-out from the
SFG photons by a harmonic-separator mirror and filters, and its
power was measured by a sensitive InGaAs detector.  In order to
demonstrate two-photon interference oscillations, a Mach-Zehnder
interferometer was placed between the last prism and the SFG
crystal.}
\end{center}
\end{figure}

Our experimental setup followed that of
\cite{Dayan@Silberberg_QPH_2004} and is depicted in Fig.
\ref{fig1}.
Basically, entangled photons generated by down-conversion of a
continuous pump-laser ($\delta_p\approx 5\:MHz$ around $532\:nm$)
in one crystal were imaged through a set of four Brewster-angle
dispersion prisms onto a second crystal, to generate the SFG
photons. This arrangement enabled a complete filtering-out of the
pump with very low loss of down-converted photons; it also enabled
a tuned compensation of dispersion (mainly of the crystals),
thereby avoiding a significant reduction of the bandwidth that was
effective for the SFG process. Finally, the spectral separation
induced by the prisms made it possible for us to turn the entire
optical layout between the crystals into a pulse-shaper.
Specifically, by using the lens after the first crystal to focus
the beam at the mid-way between the second and third prisms, we
created there a spectral Fourier plane. Then, by placing a
computer-controlled spatial light-modulator (SLM) in that plane,
we were able to introduce arbitrary phase shifts to different
spectral-components of the entangled photons
\cite{Weiner@Wullert_OL_1990}. The nonlinear crystals were
periodically-poled $KTiOPO_4$ (PPKTP) crystals, which were
temperature-controlled to obtain broadband phase matching of
$\Delta_\textsc{dc}\approx 31\:nm$ around $1064\:nm$. This
bandwidth implies $\Phi_{max}=8.2\cdot10^{12}\:s^{-1}$, which
corresponds to the classically-high power level of about $1.5\:\mu
W$. Our measurements were performed at power levels of about
$0.25\:\mu W$ ($n\approx0.16$).\

\begin{figure}[t]
\begin{center}
\includegraphics[width=8.6cm] {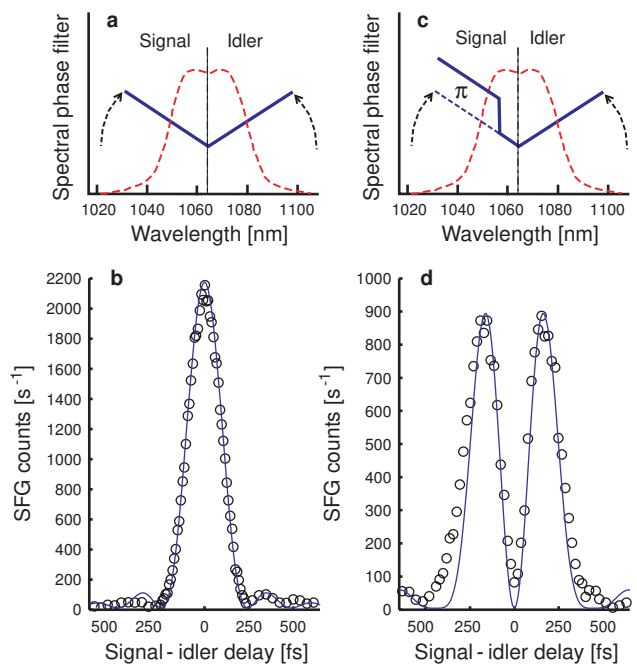}
\caption{\label{fig2} Direct observation and shaping of the
two-photon wavefunction. The relative delay between the signal and
idler photons was induces by varying the slopes of the spectral
phase filter described in (a). (b), The SFG counts (circles) and
the absolute-square of the calculated two-photon wavefunction
(line) as a function of the signal-idler delay. Shaping of the
two-photon wavefunction was performed by adding a phase
step-function in the middle of the signal spectrum, as described
in (c). The SFG measurements of the shaped wavefunction are
represented in (d)}
\end{center}
\end{figure}

To observe the two-photon wavefunction we introduced a relative
delay between the signal and idler photons of each pair, by using
the SLM to apply linear spectral-phase functions with opposite
slopes to the wavelengths below the degeneracy point of $1064\:nm$
(signal) and above it (idler), as depicted in Fig. \ref{fig2}a.
As described earlier, in this case of continuous, broadband
down-conversion, a delay has no separate effect on neither the
signal nor the idler photons. Nevertheless, as depicted in Fig.
\ref{fig2}b, the response of the SFG rate to the relative delay
between the signal and the idler photons directly reflects the
two-photon wavefunction. In order to demonstrate our ability to
coherently shape the two-photon wavefunction we applied a phase
step-function with amplitude of $\pi$ at the middle of the signal
spectrum (roughly at $1057 \: nm$, see Fig \ref{fig2}c). With
coherent pulses, this has the effect of splitting a pulse in time
to two lobes. Figure \ref{fig2}d depicts exactly such a behavior
of the two-photon wavefunction. We note that these two lobes have
opposite signs, which means there is a phase-shift between the
events where the signal photon arrives before the idler and the
events where it arrives after it. This phase shift, although
undetectable by any measurement performed on the signal (or idler)
photon itself, can be detected by interfering the SFG field with
the pump laser.\

It is important to understand under which conditions SFG can be
considered a good measure for photon coincidences. The phase
matching conditions in the nonlinear crystal dictate two relevant
bandwidths (or three, in the non-degenerate case) for the process;
one is the phase-matched bandwidth of the input (low frequency)
fields $\Delta_\textsc{lf}$, and the other is the phase-matched
bandwidth for the up-converted field $\delta_\textsc{uc}$. Loosely
speaking, for a SFG event to occur, three conditions must be
satisfied. First, the spectrum of the input photons must lie
within $\Delta_\textsc{lf}$. Second, the spectrum of the
up-converted photon must lie within $\delta_\textsc{uc}$. Third,
the incoming photons must overlap temporally to within
$\Delta_\textsc{lf}^{-1}$. Obviously, only the third condition is
equivalent to a coincidence detection. Therefore, SFG can be
considered a coincidence detector only if the entire bandwidth
$\Delta$ of the photons can be up-converted to a phase-matched
wavelength, i.e. only if:
\begin{eqnarray}
\label{cond1} \Delta_\textsc{lf}\geq\Delta \: , \ \
\delta_\textsc{uc}>2\:\Delta \:.
\end{eqnarray}
Typically, the second condition in Eq. \ref{cond1} does not hold,
since in most cases
$\delta_\textsc{uc}\ll\Delta_\textsc{lf},\Delta$, especially when
long crystals are utilized to increase the interaction efficiency.
However, this condition can be circumvented with energy entangled
photons. At low down-converted photon fluxes most of the SFG
events are the result of up-conversion of entangled pairs.
Specifically, the rate of SFG events with spontaneously
down-converted light can be approximated by:
\begin{eqnarray}
\label{R} R \propto \delta_\textsc{uc}\: n^2 + \Delta_\textsc{dc}
\: n^2 + \Delta_\textsc{dc} \: n \ .
\end{eqnarray}
As described in \cite{Dayan@Silberberg_QPH_2004}, the last term,
which exhibits a nonclassical linear intensity dependence and
therefore dominates at $n\ll 1$, is the one that results from SFG
of entangled pairs \cite{Shaped_entangled_note3}. Since the
sum-energy of such pairs is always equal to the energy of the pump
photons, for entangled-photons fluxes below $\Phi_{max}$ the
second condition of Eq. \ref{cond1} can practically be replaced
by:
\begin{eqnarray}
\label{cond2} \delta_\textsc{uc}>\delta_p \ .
\end{eqnarray}
This condition is easily met for continuous down-conversion, where
the pump is typically very narrowband, as was in our case, where
$\Delta_\textsc{lf}=\Delta=\Delta_\textsc{dc}$ (since identical
crystals were used for the down- and up-conversions), and
$\delta_\textsc{uc}\approx 0.1\:nm \gg\delta_p$.\

As is evident from Eq. \ref{R}, at high powers the contribution of
entangled pairs to the overall SFG rate becomes negligible. All
previous experiments that involved SFG with down-converted light
\cite{Abram@Dolique_PRL_1986,{Dayan@Silberberg_QPH_2003}} were
performed at extremely high powers ($\Phi \gg \Phi_{max}$). The
observation and the temporal shaping of the coherent SFG process
(denoted by the second term in Eq. \ref{R}) in these experiments
are a direct result of the fact that $\delta_\textsc{uc}$ was
orders of magnitude smaller than the bandwidth of the light, thus
violating the condition of Eq. \ref{cond1}. In such a case, many
spectral combinations of the low-frequency fields contribute to
the same narrowband up-converted field, and the resulting quantum
interference can be controlled by manipulating the spectral phase
of the incoming light, be it high-power down-converted light or
shaped pulses. The effects observed in
\cite{Abram@Dolique_PRL_1986,{Dayan@Silberberg_QPH_2003}} do not
reflect coincidences (nor the fourth-order coherence of the
light), and were demonstrated with shaped classical pulses as well
\cite{Zheng@Weiner_OL_2000}; had these experiments been performed
with extremely ultrafast detectors (or with SFG with a large
$\delta_\textsc{uc}$), the coincidence rate would have registered
mostly the regular bunching expected from thermally-distributed
light.

The complementary aspect of the time and energy entanglement
discussed earlier, is the fact that the coherence-time of the
photon pairs ($\delta_p^{-1}$) is much larger than the individual
photons' coherence-time ($\Delta_\textsc{dc}^{-1}$), allowing
two-photon interference even when the delay between the possible
paths is too large to enable one-photon interference
\cite{Franson_PRA_1991,Kwiat@Chiao_PRA_1990,Rarity@Saleh_PRL_1990,
Ou@Mandel_PRL_1990,Brendel@Martienssen_PRL_1991,Ekert@Palma_PRL_1992}.
To observe such an interference we placed a Mach-Zehnder
interferometer in the entangled-photons path after the last prism
(see Fig. \ref{fig1}). The interferometer was constructed by
rotating the polarization of the entangled photons to $45^\circ$,
utilizing the SLM to induce retardation between the vertical and
horizontal polarizations, and then rotating the polarization
$45^\circ$ back. In this configuration, the horizontal and
vertical output polarizations are essentially the two output ports
of the interferometer. In order to introduce retardations which
were beyond the range of our SLM, we placed a $1\:mm$ birefringent
Calcite crystal immediately after the SLM, thus adding a constant
retardation of about $163\:\mu m$.\

The dependence of the coincidence rate on the delay $\tau$ between
the interferometer arms is proportional to the absolute-square of
the two-photon wavefunction, and can be represented by:
\begin{equation}\label{cos}
R(\tau)\propto \left| \int g(\omega)\big( \cos\left(\omega_o \tau
\right) + \cos\left( \left(\omega-\omega_o\right) \tau \right)
\big)d\omega \right|^{2}
\end{equation}
\noindent where the symmetry of $g(\omega)$ about
$\omega_o=\omega_p/2$ was taken into account. Considering Eq.
\ref{cos}, we see that at small retardations the SFG rate is
proportional to $|(cos(\omega_o\tau)+1)|^2$, i.e. it is
quadratically proportional to the IR intensity oscillations, as
classically expected. For retardations which are beyond the
single-photon coherence-length, the IR oscillations are washed
out, and so is the second term in the integrand of Eq. \ref{cos},
for the same reasons. Thus, the SFG rate becomes proportional to
$|cos(\omega_o\tau)|^2$, which oscillates with periodicity of
$2\:\omega_o=\omega_p$.
\begin{figure}[t]
\begin{center}
\includegraphics[width=8.6cm] {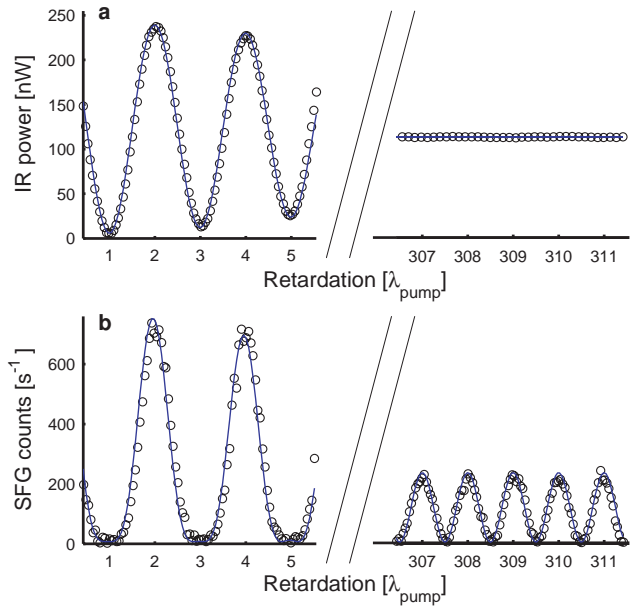}
\caption{\label{fig3} Two-photon interference of time and energy
entangled photons in a Mach-Zehnder interferometer (see Fig.
\ref{fig1}). Interference oscillations of the IR power (a) and the
SFG rate (b) as a function of the relative retardation between the
interferometer arms. The experimental SFG counts (circles) are
compared to the interference oscillations of the two-photon
wavefunction, calculated according to Eq. \ref{cos} (line).}
\end{center}
\end{figure}
The two-photon interference oscillations maintain their high
visibility due to the long coherence length of the entangled
pairs, which enables interference between the events at which a
photon pair travelled unseparated along one interferometer arm, or
the other. However, whenever such oscillations were detected with
photoelectric coincidences
\cite{{Kwiat@Chiao_PRA_1990},Brendel@Martienssen_PRL_1991}, the
visibility dropped below $50\%$ when the relative delay ranged
between the coherence time of the photons ($\sim 100\:fs$) to the
$4$ orders of magnitude longer temporal resolution of the
coincidence circuits ($\sim 1\:ns$). In contrast, the temporal
resolution of the SFG process is equal to the coherence time of
the photons since $\Delta_\textsc{lf}=\Delta_\textsc{dc}$. Thus,
it performs as a 'background-free' coincidence detector,
inherently rejecting non-coinciding photons. As can be seen in
Fig. \ref{fig3}, the SFG counts follow the interference
oscillations of the two-photon wavefunction throughout the entire
range of delays, demonstrating visibility of $94\%\pm4\%$ at a
relative delay of $\sim550\:fs$, where the single-photon
interference completely dies-out. Note that $50\%$ is the
classical limit in the absence of first-order interference
\cite{{Kwiat@Chiao_PRA_1990},
{Su@Wodkiewicz_PRA_1991},Brendel@Martienssen_PRL_1991}.\

To conclude, we demonstrated how pulse-shaping techniques and SFG
can be exploited to precisely control and measure the temporal
properties of entangled photon-pairs. Although we emphasized the
ability of SFG to directly reflect the shape of the two-photon
wavefunction, spectral-phase manipulations can be observed and
utilized in schemes based on coincidence-measurements as well
\cite{Hong@Mandel_PRL_1987}. We believe both tools have potential
applications in quantum measurement and quantum information
science.


\begin{thebibliography}{24}
\bibitem[{Sha({\natexlab{b}})}]{Shaped_entangled_note1}
\bibinfo{note}{The bandwidth $\delta_p$ was taken as full-width at $1/e$. Note
  that this analysis does not hold for type-II down-conversion, where the
  wavefunction is broadened due to the group-velocity difference between the
  photons.}
\bibitem[{Sha({\natexlab{c}})}]{Shaped_entangled_note2}
\bibinfo{note}{This description assumes the extreme case, where the pump is
  almost as broadband as the down-converted pulses.}
\bibitem[{Sha({\natexlab{d}})}]{Shaped_entangled_note3}
\bibinfo{note}{Nonclassicality is exhibited by nondegenerate entangled photons
at $n\ll1$ since
$|\textrm{g}^{(2)}_{12}(0)|^2>\textrm{g}^{(2)}_{11}(0)\:\textrm{g}^{(2)}_{22}(0)$,
 violating the Cauchy-Schwartz inequality for classical
fields.}
\end{thebibliography}

\end{document}